# Pressure-Tunable Ambipolar Conduction and Hysteresis in Ultrathin Palladium Diselenide Field Effect Transistors

*Antonio Di Bartolomeo[1,2,3,\*], Aniello Pelella[1], Xiaowei Liu[4], Feng Miao[4,\*], Maurizio Passacantando[5], Filippo Giubileo[3], Alessandro Grillo,[1], Laura Iemmo[1,2,3], Francesca Urban[1,2,3] and Shi-Jun Liang[4]*

[1]Physics Department "E. R. Caianiello", University of Salerno, via Giovanni Paolo II n. 132, Fisciano 84084, Italy

[2]Interdepartmental Centre NanoMates, University of Salerno, via Giovanni Paolo II n. 132, Fisciano 84084, Italy

[3]CNR-SPIN Salerno, via Giovanni Paolo II n. 132, Fisciano 84084, Italy

[4]National Laboratory of Solid State Microstructures, School of Physics, Collaborative Innovation Center of Advanced Microstructures, Nanjing University, Nanjing, 210093, China.

[5]Department of Physical and Chemical Science, University of L'Aquila, and CNR-SPIN L'Aquila, via Vetoio, Coppito 67100, L'Aquila, Italy

\*E-mail: adibartolomeo@unisa.it, miao@nju.edu.cn

**Keywords:** palladium diselenide, field effect transistors, hysteresis, electron irradiation, ambipolar conduction

**Abstract:**

A few-layer palladium diselenide ($PdSe_2$) field effect transistor is studied under external stimuli such as electrical and optical fields, electron irradiation and gas pressure. We observe ambipolar conduction and hysteresis in the transfer curves of the $PdSe_2$ material unprotected and as-exfoliated. We tune the ambipolar conduction and its hysteretic behavior in the air and pure nitrogen environments. The prevailing p-type transport observed at room pressure is reversibly turned into a dominant n-type conduction by reducing the pressure, which can simultaneously suppress the hysteresis. The pressure control can be exploited to symmetrize and stabilize the transfer characteristic of the device as required in high-performance logic circuits. The transistor is immune from short channel effects but is affected by trap states with characteristic times in the order of minutes. The channel conductance, dramatically reduced by the electron irradiation during scanning electron microscope imaging, is restored after several minutes anneal at room temperature. The work paves the way toward the exploitation of PdSe$_2$ in electronic devices by providing an experiment-based and deeper understanding of charge transport in PdSe$_2$ transistors subjected to electrical stress and other external agents.

**Introduction**

The relentless search for new two-dimensional (2D) layered materials[1–6] beyond graphene[7–11] has lately identified palladium diselenide (PdSe$_2$) as a new promising candidate for next-generation electronic and optoelectronic applications.[12,13] PdSe$_2$ is the first discovered noble transition-metal dichalcogenide (TMDC) with 2D pentagonal structure.[14] Monolayer PdSe$_2$ has a puckered configuration with the Pd atoms in the middle covalently bonded to four Se atoms, two of which are located respectively in the top and in the bottom part (Figure 1(a)). Two neighboring Se atoms in the puckered morphology form a covalent Se−Se bond with ∼1.6 Å puckering distance.[14] Compared to other TMDCs,[15–20] PdSe$_2$ has a higher stability in air and is an indirect semiconductor with bandgap from 1.3 eV for the monolayer to 100 meV or less for the bulk. [12] Such a high bandgap tunability is one of the most remarkable properties of PdSe$_2$ that has no equals in other 2D semiconducting materials. This important peculiarity of PdSe$_2$ could enhance its light absorption capability.[21] The practical application of PdSe$_2$ can be further diversified by the strong spin-orbit coupling and the tunable topological quantum phase transitions,[22,23] as well as by the induced ferromagnetism with Curie temperature beyond room temperature.[24,25] Hence, 2D pentagonal PdSe$_2$ is very promising for the design of new functionalities in higher-performance electronic devices that combine the charge, spin, and other degrees of freedom resulting from the low symmetry.

The core of electronic devices applications depends on the transfer characteristics of PdSe$_2$-based field effect transistors (FETs), in which PdSe$_2$ offers ambipolar behavior, good on/off ratio and high electron and hole mobility. An hysteresis usually occurs in the ambipolar transfer curve of as-fabricated PdSe$_2$ FETs,[26] which is claimed to be caused by process residues and absorbates, but can be reduced by vacuum annealing. Hysteresis has been reported in FETs based on nanotubes,[27,28] graphene[29,30] and TMDCs.[31] In general, the charge transfer, charge trapping or charge polarization are proposed to interpret this phenomenon. However, these mechanisms are still under debate. For example, in the MoS$_2$-based field effect transistors, charge traps may arise from the trapping center at MoS$_2$/SiO$_2$ interface,[32,33] absorbates on the MoS$_2$ channel,[34,35] or intrinsic sulfur vacancies or other defects.[31,36,37] It is highly desirable to have a comprehensive understanding of the hysteresis and to achieve a good control of the phenomenon, so that it can be either eliminated from transfer curves to avoid threshold voltage instability or conveniently exploited, for instance, into memory devices.[38–40]

It has also been found that the semiconducting phase of few-layer PdSe$_2$ can be changed to a semimetallic phase with an out-of-plane electric field.[14,26] Similarly, pressure can be used to mechanically tune lattice constant of PdSe$_2$ to achieve the modulation of interlayer coupling and electronic band structures.[41,42] When the pressure exceeds 3 GPa semiconducting to metal phase transition occurs in single crystal PdSe$_2$. Further increase in the pressure over 6 GPa transfers the structural phase of PdSe$_2$ to the pyrite phase, where the superconductivity emerges with critical temperature rapidly increasing in correlation with a weakening of the Se-Se bonds.[41,42] Combining pressure and electric field control results in mechanical and electrostatic tuning of the crystalline structural and electronic properties and enables tunable bipolar behavior and hysteresis. This can make PdSe$_2$ suitable for potential applications in nanoelectromechanical devices and future complementary logic electronics.

In this paper, we exfoliate bulk crystals and fabricate back-gate field effect transistors to investigate several transport properties in few-layer PdSe$_2$ under external stimuli such as electrical and optical field, electron irradiation, gas exposure, and so forth. We present the transistor electrical characterization, with focus on the effects of the drain and gate voltage stress. We show that unprotected and as-fabricated PdSe$_2$ devices with Ti contacts exhibit ambipolar conduction with electron (hole) mobility up to 4 (3) cm$^2$V$^{-1}$s$^{-1}$, although measured a few months after the production. We study the gate hysteresis and the trap dynamics and show how the exposure to electron beam irradiation, usually performed for imaging purposes, can have a temporary dramatic effect on the device performance. Importantly, we show that the control of pressure in air or pure-nitrogen ambient is a good knob to balance between the electron or hole conduction and to suppress the hysteresis. Hence, it can be used to set the symmetry of the transfer characteristics and to stabilize the device. These unexplored and important features of the electric transport in PdSe$_2$ in our work are essential for the practical exploitation of this new material.

**Results and discussion**

A scanning electron microscope (SEM) top-view of the device and its schematic cross-section are shown in Figures 1(b) and its inset. The atomic force microscope (AFM) image in Figure 1(c) reveals a thickness of 15 nm for the exfoliated flake, which corresponds to about 25 layers.[14] The energy dispersive X-ray spectrum and the X-ray diffraction pattern in Figures 1(d) and 1(e), respectively, indicate a Pd:Se atomic ratio close to 1:2 and a low-defect crystalline structure. The Raman spectrum of the flake (Figure 1(f)) displays five main distinct peaks slightly shifted with respect to theoretical peaks of bulk PdSe$_2$, consistent with the estimated number of layers.[14,21,26]

The electrical characterization of the PdSe$_2$ transistor is summarized in Figure 2. The output characteristics, i.e. the $I_{ds}$ drain-source current as a function of the $V_{ds}$ drain-source voltage with the $V_{gs}$ gate-source voltage as the control parameter, shown in Figure 2(a), are symmetric and linear revealing an ohmic behavior over the considered bias range. The linear $I_{ds} - V_{ds}$ behavior is preserved under the application of the gate voltage, which only affects the overall drain-source conductance. The modulation of the channel current is further investigated with the transfer characteristic, $I_{ds} - V_{gs}$ curve at given $V_{ds}$, measured over a loop of the gate voltage and displayed both on logarithmic and linear scale in Figure 2(b). The ~25 on/off ratio is a consequence of the number of layers, which imply a reduced bandgap and then a limited gate control.

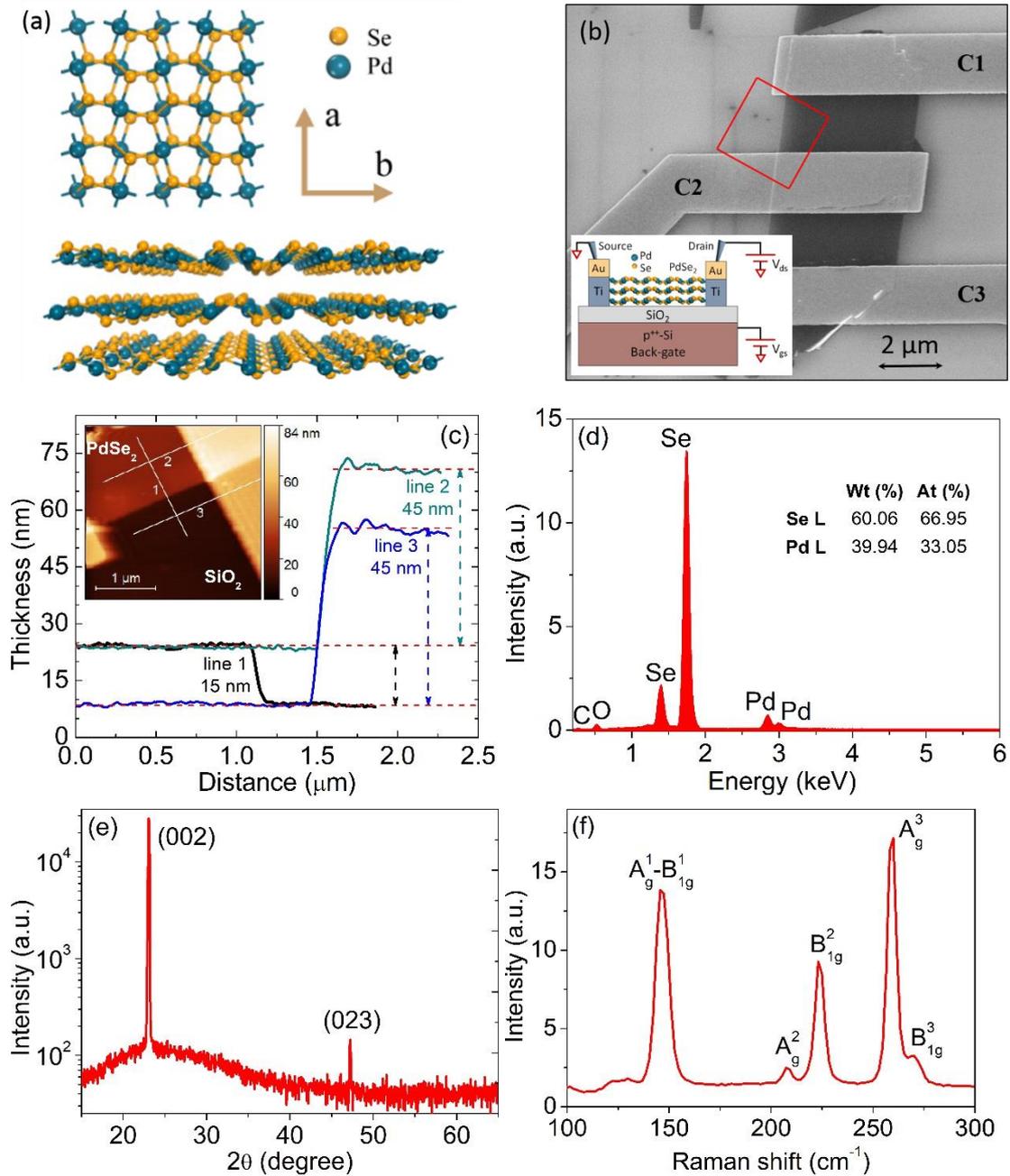

Figure 1: (a) Top and side view of the puckered pentagonal configuration of thin film $PdSe_2$. (b) Scanning electron microscope image of the $PdSe_2$ flake with 5 nm Ti/40 nm Au metal contacts and (inset) schematic of the back-gate transistor fabricated with it (not-on-scale). (c) Atomic force microscope image of the part of the device in the red box of (b) and height profile along the lines marked as 1, 2 and 3 in the inset. The flake has a step height of 15 nm from the $SiO_2$ floor, that corresponds to ~25 $PdSe_2$ layers (see black line). (d) Energy dispersive X-ray spectrum, (e) X-ray diffraction pattern and (f) Raman spectrum of the flake.

Differently from similar FETs based on TMDCs like MoS$_2$[33,43] or WSe$_2$,[16,18] the PdSe$_2$ transistor exhibits a clear ambipolar behavior with a slight electron-hole asymmetry, and prevailing n-type conduction in high vacuum. The behavior mimics that observed in graphene transistors,[44–48] with a minimum conduction point ($V_{gs}^{min}, I_{ds}^{min}$), separating the n-type conduction (for $V_{gs} > V_{gs}^{min}$) from the p-type one (for $V_{gs} < V_{gs}^{min}$). The minimum point is affected by the biasing history and is different during the reverse and forward sweeps. A prevailing n-type behavior and a wide hysteresis between the two sweeps are also evident in the transfer characteristic.

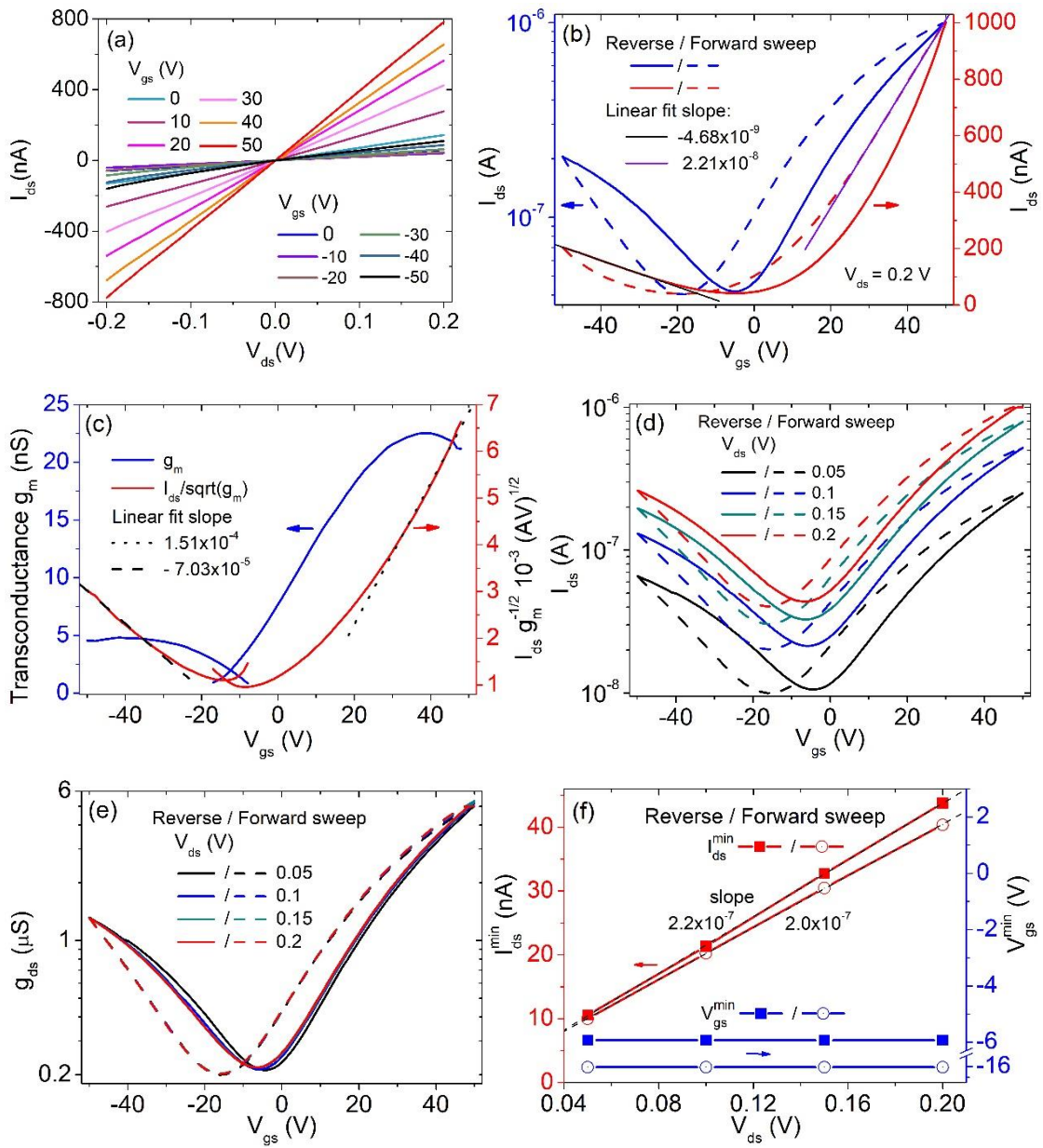

Figure 2: Output (a) and transfer (b) characteristics of the PdSe$_2$ transistor measured at pressure

<$10^{-6}$ Torr. The fitting straight lines shown in (b) are used to evaluate the field effect mobility. Reverse and forward refer to $V_{gs}$ swept from 50 to -50V and from -50 to 50V, respectively. (c) Transconductance $g_m$ and $\frac{I_{ds}}{\sqrt{g_m}}$ ratio vs gate voltage, with linear fits to extract the mobility deduced of the effect of the contact-resistance. PdSe$_2$ FET current (d) and channel conductance (e) versus $V_{gs}$ for different drain biases. (f) Gate voltage and current ($V_{gs}^{min}$ and $I_{ds}^{min}$) at the conductance minimum for the reverse and forward sweeps.

In the linear region, the FET drain-source current can be expressed as

$$I_{ds} = \frac{W}{L}\mu_{FE}C_{ox}(V_{gs} - V_{th})^\alpha V_{ds}, \quad (1)$$

where W and L are the channel width and length, $\mu_{FE}$ is the field effect mobility, $C_{ox} = 1.15 \cdot 10^{-8} \frac{F}{cm^2}$ is the capacitance per unit area of the 300 nm SiO$_2$ gate dielectric, $V_{th}$ is the threshold voltage and $\alpha \geq 1$ is a dimensionless parameter which accounts for a possible $V_{gs}$-dependence of the mobility.[31,49,50] According to Eq. (1), when the $I_{ds} - V_{gs}$ curve is linear, $\alpha = 1$, and the mobility can be obtained as

$$\mu_{FE} = \frac{L}{W}\frac{1}{C_{ox}}\frac{1}{V_{ds}}g_m, \quad (2)$$

where $g_m = \frac{\partial I_{ds}}{\partial V_{gs}}\bigg|_{V_{ds}=const}$ is the FET transconductance. Using the data and the slopes of the linear fittings of Figure 1(b), Eq. (2) yields a hole and electron field effect mobility of ~0.86 cm$^2$V$^{-1}$s$^{-1}$ and ~4.1 cm$^2$V$^{-1}$s$^{-1}$, respectively.

In two-probe measurements, $\mu_{FE}$ can be negatively affected by a high contact resistance $R_c$.[18,51] However, the effect of $R_c$ can be taken into account by replacing $V_{ds}$ with $V_{ds} - R_c I_{ds}$ in Eq. (1). Assuming that the contact resistance is independent of $V_{gs}$, as corroborated by the preserved linearity of the output curves with increasing $V_{gs}$, it can be easily shown[52] that

$$\frac{I_{ds}}{\sqrt{g_m}} = \sqrt{\frac{W}{L}\mu_{FE}C_{ox}V_{ds}}\,(V_{gs} - V_{th}). \quad (3)$$

Eq. (3) is used to extract the mobility, deduced by the effect of the contact resistance, from the fit of the $\frac{I_{ds}}{\sqrt{g_m}}$ vs $V_{gs}$ plot shown in Figure 2(c). The $R_c$-corrected hole and electron mobilities (0.90 cm$^2$V$^{-1}$s$^{-1}$ and 4.2 cm$^2$V$^{-1}$s$^{-1}$) do not show a significant improvement, confirming a negligible effect of the contacts. Although higher electron (up to 216 cm$^2$V$^{-1}$s$^{-1}$)[26] and hole mobilities (up to ~20 cm$^2$V$^{-1}$s$^{-1}$)[14] have been achieved in few-layer PdSe$_2$ transistors, values comparable to those here quoted have been reported for PdSe$_2$ ultra-thin films of similar thicknesses. Indeed, the mobility decreases rapidly with the number of layers after reaching a peak at 7-10 layers, a behavior found also in other puckered materials such as phosphorene.[14,53] We note also that the mobility is here measured on a device without any channel material treatment or optimization, and its enhancement with further device engineering can realistically be envisioned.

The effect of the drain bias on $I_{ds}$ is studied in Figures 2(d)-(f). The increasing $V_{ds}$ results in a growing current without an appreciable effect on the transistor conductance ($g_{ds} = \frac{dI_{ds}}{dV_{ds}}$), as displayed in Figure 2(e). This observation confirms the immunity of the device from short channel effects like the drain-induced barrier lowering, within the explored range. More importantly, $V_{ds}$ does not affect $V_{gs}^{min}$, neither for the reverse nor for the forward sweep (Figure 2(f)), implying a negligible effect of the drain bias also on the transistor hysteresis. As the hysteresis is due to charge trapping in localized trap states,[27,31,34,54–57] this finding excludes the trapping modulation by lateral electric field at the interface with SiO$_2$, recently reported for instance in back-gate MoS$_2$ field-effect transistors.[58]

Figure 3 shows two important features of the gate-induced hysteresis. By quantifying the hysteresis though its width $H_w$ defined as the difference of the gate voltages corresponding to a current $I_{ds}$ =50 nA, we observe that $H_w$ increases quadratically with the $V_{gs}$ sweeping range, and is an exponentially-growing function ($a - b \cdot e^{-x}$) of the $V_{gs}$ sweeping time. As already mentioned, the gate-induced hysteresis in FETs is caused by charge transfer from/to intrinsic and extrinsic trap states. Intrinsic traps correspond to PdSe$_2$ crystal defects such as vacancies or grain boundaries,[59,60] while extrinsic traps are related to adsorbates from the environmental exposure, such as H$_2$O and O$_2$ molecules, or to residues from the fabrication process.[31,34,61,62] Water and oxygen, although identified as major contributors to the hysteresis of 2D-material FETs,[63,64] are expected to be uninfluential in the measurements of Figure 3 because of their desorption after several-hour annealing in high vacuum. Then, traps at the PdSe$_2$/SiO$_2$ interface and intrinsic PdSe$_2$ defects or border traps in SiO$_2$ as well as mobile charge in the SiO$_2$ layer are under the spot as possible cause of hysteresis.[65–67]

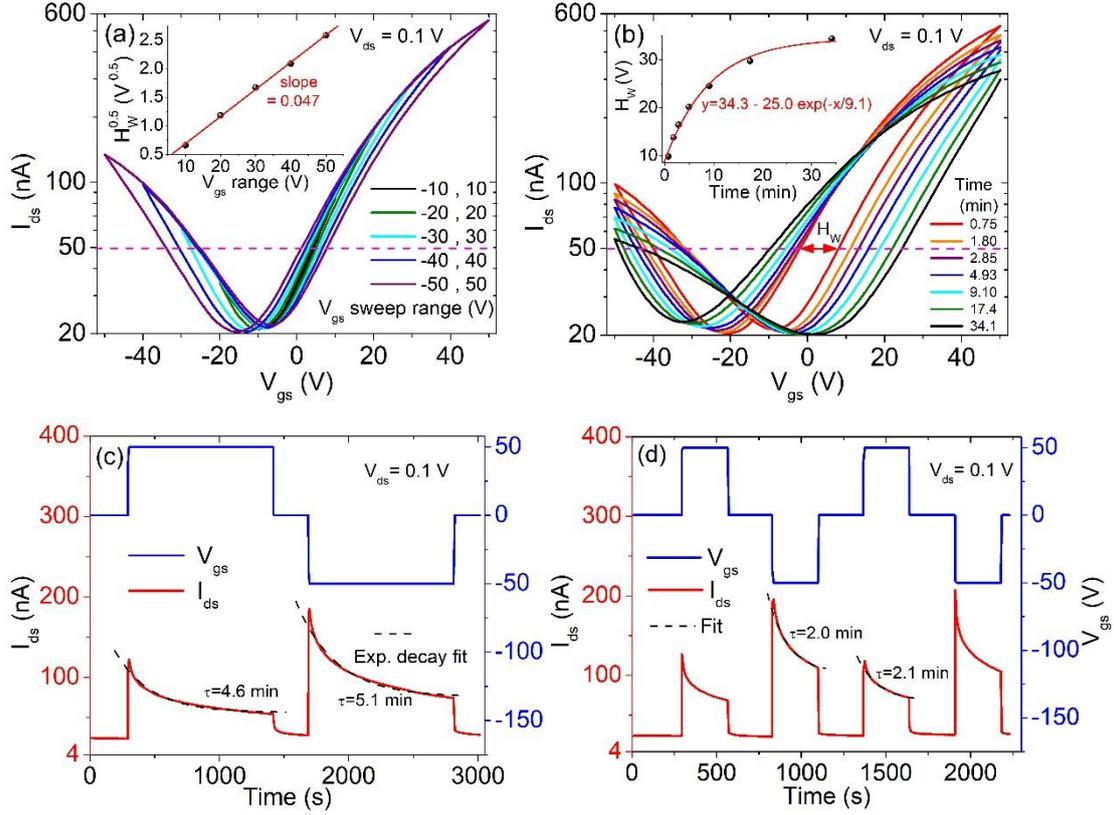

Figure 3: Transfer characteristics of the PdSe$_2$ transistor for different $V_{gs}$ (a) sweeping range and (b) sweeping rate. The inset of (a) reports the square root of the hysteresis width, $\sqrt{H_W}$, as a function of $V_{gs}$ sweeping range, while the inset of (b) shows the hysteresis width, $H_W$ as a function of the sweeping time and the exponential-decay fitting. $H_W$ is estimated on the n-branch of the transfer characteristics as the difference of the $V_{gs}$ values corresponding to $I_{ds}$ = 50 nA. Drain-source current of the device subjected to $V_{gs} = \pm 50$ V pulses of length (c) 20 min and (d) 3 min, respectively.

The quadratic $H_w - V_{gs}$ behavior, common to graphene back-gate transistors,[29] excludes a dominant role of mobile-charge motion inside the SiO$_2$ dielectric. In fact, the latter would make $H_w$ a weaker function of $V_{gs}$ as the involved total charge would be almost constant during the sweep. The RC-exponential growth observed in $H_W$ as a function of the sweeping time, with a single time constant $\tau = RC \approx 9$ min, also diminishes the role of PdSe$_2$/SiO$_2$ interface traps, which are well-known fast states (in the milliseconds range),[68] and corroborates the hypothesis of slow trap states related to either PdSe$_2$ or SiO$_2$ defects. While the nature of intrinsic traps in PdSe$_2$ (perhaps non the most important ones, considering the high crystallinity of the sample) is still unclear, the slow border traps in SiO$_2$, are attributed to trivalent silicon dangling bonds or hydrogenic defects.[69] Using the RC time constant, we can make an estimation of the involved capacitance considering that R is the inverse of the transconductance, which is ~10 nS (Figures 2(c)). This corresponds to a capacitance of ~5 µF, which is far higher than that of the gate oxide of the device, in the order of the pF, implying that the trap-related capacitance is the dominant one. The same conclusion can be reached considering the sub-threshold swing $SS$, that is the gate voltage change corresponding to one-decade increase of the transistor current, that is expressed in terms of the trap ($C_T$) and channel depletion layer ($C_{DL}$) capacitances per unit-area as

$$SS = \frac{dV_{gs}}{d\log I_{ds}} \approx \ln(10)\frac{kT}{q}\left(1 + \frac{C_T+C_{DL}}{C_{ox}}\right) \quad (4)$$

(here, $k$ is the Boltzmann constant, $T$ is the temperature). Assuming $C_{DL}$ negligible compared with $C_T$ (a reasonable assumption considering the low modulation of the current), the relatively high $SS \sim 30$ V/decade obtained from Figure 2(b) or 3(b) yields the consistent trap capacitance, $C_T$= 5.7 µF, and a density of trap states $D_T = \frac{C_T}{q^2} \approx 3.5 \cdot 10^{13}$ cm$^{-2}$ eV$^{-1}$.

A further confirmation of the several-minute time constants attributed to the traps is provided by the transient behavior of the device, investigated through a series of $V_{gs} = \pm 50$ V gate pulses in high vacuum, shown in Figures 3(c) and 3(d). The trapping of charge is here seen as a reduction of the device current while the gate pulse is in the high positive or negative state. The RC delay from the $V_{gs}$ pulses, consistent with that from $H_W$ transient, confirms the slow trap states. Furthermore, the independence of the time constant of the gate polarity indicates similar electron and hole capture/emission times.

The good stability and robustness of the device is reflected in the preservation of the current level and ambipolar behavior in the typical switching sequence displayed in Figure 3(d).

Charge trapping at the PdSe$_2$ defects and into the SiO$_2$ topmost layers is also responsible for the device behavior under electron irradiation. Figure 4(a) compares the transfer characteristics of the transistor before and after the irradiation consequent to SEM imaging with 10 keV electron beam, corresponding to a fluence of about 4 electrons/nm$^2$. The electron beam irradiation has a dramatic effect, resulting in about an order-of-magnitude reduction of the channel conductance and a shift of the conductance minimum. The reverse and forward sweep curve moves rightwards and leftwards, respectively. Figure 4(a) also shows that, after irradiation, the transfer characteristic slowly recovers approaching the initial state in a time of the order of the hours. The electron beam generates electron-hole pairs in both the channel layer and the underlying dielectric and promotes the formation of defects.[43,47,70–72] This favors charge trapping and degrades the PdSe$_2$ electrical conductivity. The 10 keV electrons are absorbed mostly in the SiO$_2$ layer, where they create a pile-up of negative charge, which screens the gate field and affects the channel carrier conductivity. For positive $V_{gs}$, the electrons injected in the SiO$_2$ are attracted towards the Si-gate and oppose the positive gate voltage. As a consequence, a higher gate voltage is needed to start the n-branch (right shift of $V_{gs}^{min}$); conversely, for negative $V_{gs}$, the electrons from the beam are pushed toward the PdSe$_2$ channel and contribute to its n-doping, such that a higher negative $V_{gs}$ is needed to initiate the p-branch (left shift of $V_{gs}^{min}$). The effect of the electron beam vanishes mostly with time rather than with the sweep repetition.

Indeed, Figures 4(b) and 4(c) show that the restoring of $V_{gs}^{min}$ and $I_{ds}^{min}$ after irradiation, at room temperature, follows an exponential decay law with time constant of several minutes, consistent with the previously estimated trap time-constant. Such observation confirms the key role played by the charge transfer and trapping between the PdSe$_2$ channel and the underlying SiO$_2$ gate dielectric.

Figure 4(d) shows a similar exponential recovery for the (n-branch) hysteresis and for the electron and hole mobilities. Such a behavior is easily understood considering that the slow diffusion or drift of the electrons from the beam, piled up in SiO$_2$, to the PdSe$_2$ channel (or the Si gate) simultaneously reduces the hysteresis and the Coulomb scattering which limits the carrier mobility. However, we note that the initial state is not completely restored, probably indicating some permanent radiation-induced channel damage.

We also checked the effect of optical irradiation but the device did not show any appreciable photoresponse, either in prevailing n- or p-mode. Figure 4(e) shows unchanged transfer characteristics upon illumination by the 880 nm LED used for optical imaging in the SEM chamber. Analogously, Figure 4(f) demonstrates that a strong white LED light or the laboratory illumination do not appreciably affect the typical increasing p-type current during air exposure.

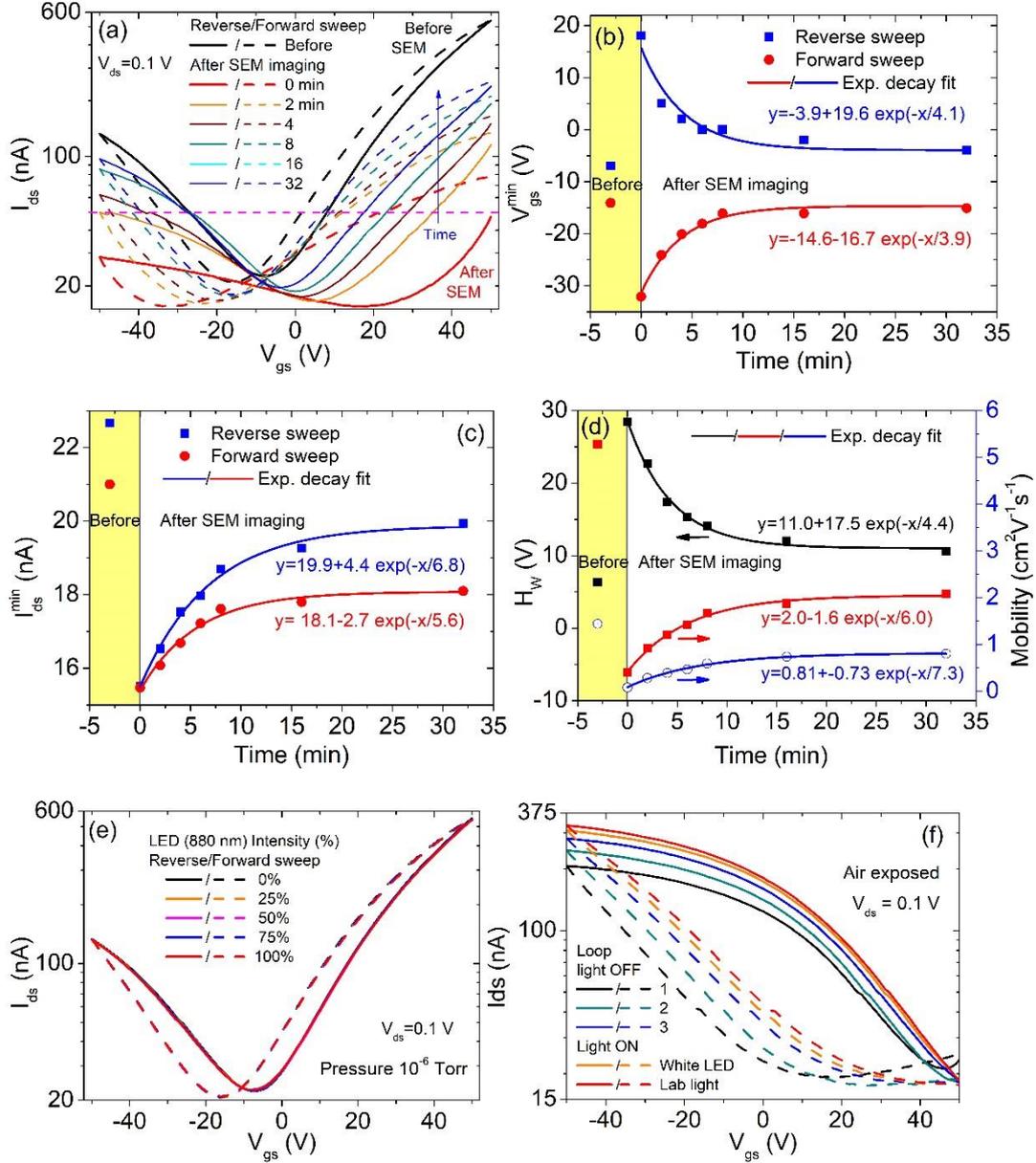

Figure 4: Transfer characteristics before and after SEM imaging by means of an electron beam with energy and current of 10 keV and 6 pA, respectively, and scanning time of 31 s corresponding to a fluence of 4 electrons/nm$^2$ (for clarity, only a subset of the measured curves is shown here). Time evolution of (b) $V_{gs}^{min}$, (c) $I_{ds}^{min}$ and (d) hysteresis width plus electron-hole mobility, after irradiation. Photo response of the device (e) in high vacuum illuminated with the 880 nm LED of the SEM chamber at progressive light intensities up to the 265 µWcm$^{-2}$ and (f) in air under white LED and laboratory illumination.

The control of the n- and p-type behavior, as well as of the hysteresis, is an important prerequisite for the transistor exploitation in practical circuits, with ambipolar symmetry, high on/off ratio and low hysteresis highly desirable for stable low-power logic applications. Here, we show that exposure to air or nitrogen offers an easy knob to balance between n- or p-type conduction and to reduce the hysteresis. Favored by chalcogen vacancies and due to their high electronegativity, adsorbed $O_2$, $N_2$ and $H_2O$ molecules capture electrons and induce a p-type doping in TMDC materials.[73–75] Indeed, Figure 5(a) shows that increasing pressure by air in a time of a few minutes has a dramatic effect on the $PdSe_2$ transistor. The increasing pressure gradually reduces the n-type conduction in favor of the p-type one and transforms the device from a prevailing n-type to a prevailing p-type transistor. In particular, the exposure to air for 10 min changes the transistor in a p-type depletion mode device. More importantly, the swapping from n- to p-type conduction is reversible, as demonstrated in Figure 5(b), where the opposite trend from p- to n-type conduction is measured for the decreasing pressure. In particular, the $10^{-6}$ Torr vacuum after about 10 hours resets the device close to its initial n-type state. Remarkably, Figure 5(c) shows that lowering the pressure reduces the hysteresis, both on the n- and p-branches. This effect, which is likely due to the gradual desorption of the adsorbates, provide a viable approach to the hysteresis control. In particular, it can be noted that a pressure below $10^{-4}$ Torr is as effective in reducing the hysteresis as is a pressure approaching the atmospheric one in increasing it.

The electron and hole mobilities, obtained from both pressure cycles, show a decreasing trend with increasing pressure (Figure 5(d)), as reported also in graphene transistors.[76] Figure 5(e) shows a dependence of $I_{ds}^{min}$ on the pressure. While the decreasing current with increasing pressure is easily understood as the effect of decreasing mobility caused by adsorbate-induced scattering, the crossover occurring around $10^{-4}$ Torr and the subsequent smoother increasing current with increasing pressure is non-trivial. Such increase goes against the decreasing mobility and could be related to a slight variation of the material bandgap caused by the adsorbates.[77] A lower bandgap could results in increased carrier density mainly for the reduced ionization energy of the trap states which would increase their contribution to the free carrier density.

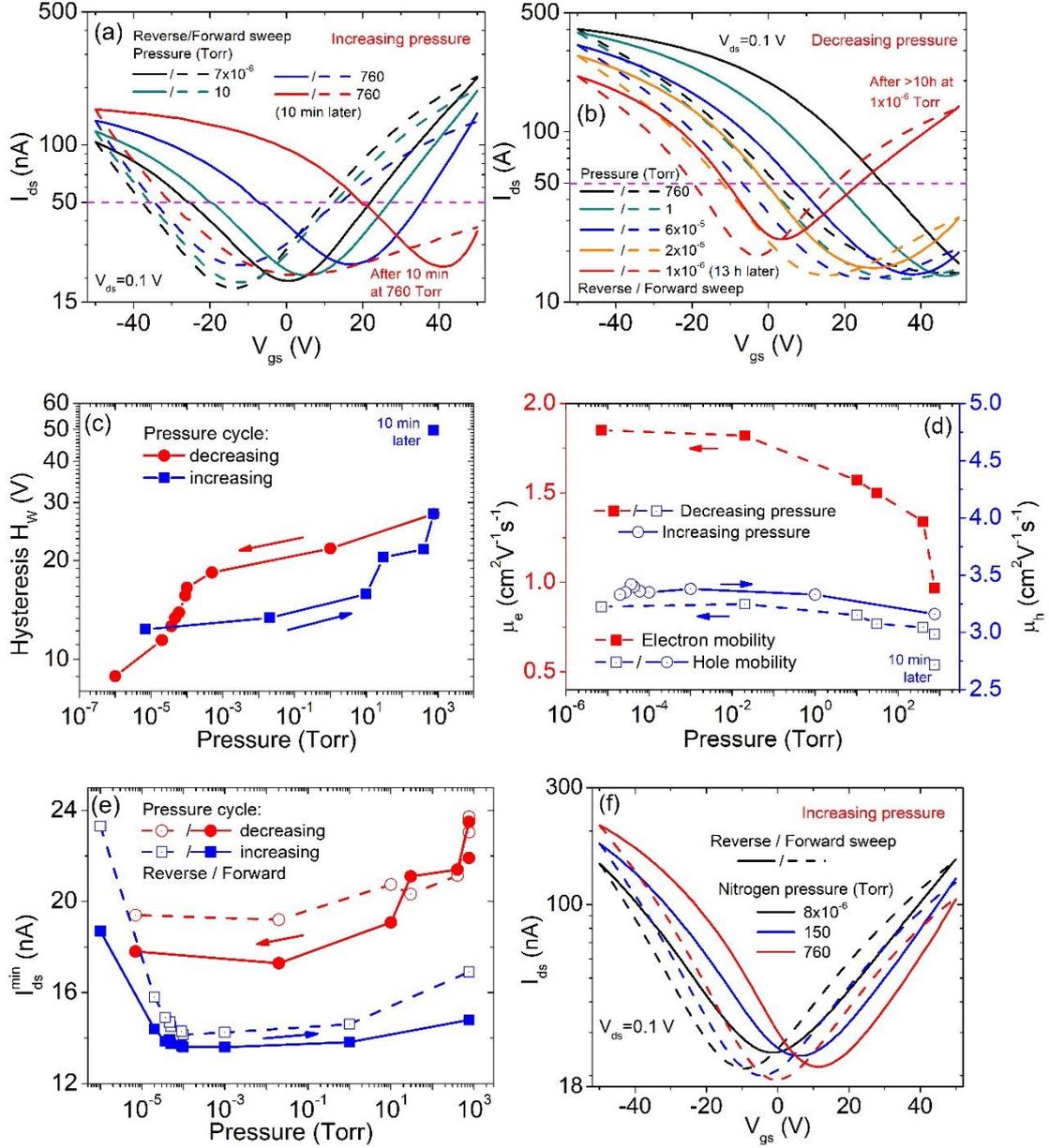

Figure 5: Effects of pressure on the transfer characteristics of the PdSe$_2$ device (a) for increasing and (b) deceasing pressure (for clarity, only a subset of the measured curves is shown here). Hysteresis (c), electron-hole mobility (d) and $I_{ds}^{min}$ (e) as a function of pressure. (f) Effects of pure nitrogen pressure on the transfer characteristics.

The enhancing effect of the pressure on the conductance and the carrier concentration has been measured for few-layer MoS$_2$ subjected to the high hydrostatic pressure, in the order of GPa, applied by means of diamond anvil cell.[78] A clear change of the transport properties of a single crystal PdSe$_2$ from the semiconducting to the metal phase occurring without a structural phase transition under the application of a high pressure of 3 GPa has been observed.[41] Moreover, an enhanced conductivity and carrier density, associated to a reduction of mobility, similar to what we have measured, has been predicted by DFT calculations in multilayer $WS_2$ subjected to high pressure, in the order of GPa, as the effect of band gap closure due to pressure-increased interlayer and sulphur-sulphur interaction.[79] The data of Figure 5(d) and 5(e) suggest that an effect of the pressure on the transport properties of multi-layer $PdSe_2$ is detectable at far lower pressures, although of not directly related to a mechanical deformation.

Inert gases are easier to handle and are suitable for pressure control. An experiment with the pressure controlled in pure nitrogen ambient, is reported in Figure 5(f), showing that $N_2$ (whose electronegativity is close to that of oxygen, 3.0 *vs* 3.5 eV) can control the n- to p-type conduction conversion similarly to air. Then, the ambipolar characteristics of few-layer $PdSe_2$ can be strongly modulated by air or nitrogen. A similar behavior has been observed in transistors with few-layer black-phosphorus exposed to oxygen.[80,81]

We highlight that these distinct pressure-tunable characteristics of $PdSe_2$ hold promises for the future development of new pressure modulated electronic devices and pressure sensors with ultra-wide dynamic range.

**Conclusions**

We have fabricated back-gate field effect transistors with exfoliated few-layer $PdSe_2$ and studied their electric behavior under several environmental conditions and external stimuli. The device exhibits an ambipolar behavior that is strongly sensitive to electrical stress, electron irradiation and pressure. We examined the dramatic effect of electron irradiation on the hysteresis and its recovery process, which are correlated to the slow trap states in the $PdSe_2$ and $SiO_2$. We have demonstrated that the control of pressure in air or pure nitrogen environment is an effective knob to switch between n- and p-type conduction and reduce the hysteresis in the transfer characteristics. This study provides new understanding and experimental evidence of the behavior of few-layer $PdSe_2$ as the channel of field effect transistors and shows the great potential of $PdSe_2$ for the development of electronic logic devices and of pressure sensors with ultra-wide range.

**Experimental session**

The flakes were prepared from bulk PdSe$_2$ single crystals using the standard mechanical exfoliation method by adhesive tape. The exfoliation is facilitated by the low interlayer binding energy of ~0.35 J/m$^2$ of PdSe$_2$ (corresponding to ~ 62 meV/atom), which is smaller than that of graphite (0.37 J/m$^2$).[82] The flakes were transferred to degenerately doped p-type silicon substrates, covered with 300-nm-thick SiO$_2$, on which they were localized and identified using optical microscopy. Standard electron-beam lithography followed by metal electron-beam evaporation were carried out to deposit 5 nm Ti/40 nm Au metal contacts. The schematic and a SEM top-view of the device here studied are shown in Figures 1(b) and its inset, respectively. The AFM image of Figure 1(c) reveals a thickness of 15 nm for the flake, which corresponds to about 25 layers (assuming a thickness of 0.6 nm for a single layer).[14]

The chemical composition of the flake was measured by Hitachi S-3400 N II SEM energy dispersive X-ray spectroscopy (EDXS), and the Pd:Se atomic ratio is close to 1:2 (Figure 1(d)). The crystalline phase of the PdSe$_2$ flakes was characterized by the Bruker D8 Discover X-ray diffractometer with Cu $K_\alpha$ radiation ($\lambda$ = 1.5406 Å) operating in Bragg-Brentano mode.

The diffraction pattern (Figure 1(e)) of the sample confirms a layered PdSe$_2$ single crystal as shown by the presence of the characteristic peak at (002) that indicates a strong orientation along the *c* axis due to the layered crystal structure along the *c* axis (the unit cell is orthorhombic with space group *Pbca*). Moreover, the crystalline peak (023), detected in the XRD pattern, can be attributed to some exfoliated flakes not oriented respect to the plane substrate.

The Raman spectrum (Figure 1(f)) was measured under excitation line of 514 nm by Renishaw inVia Raman microscope H54304 and displays five distinct peaks typical of bulk PdSe$_2$. There peaks correspond to three A$_{1g}$ and three B$_{1g}$ modes, with $A_{1g}^1$ and $B_{1g}^1$ very close and barely distinguishable. The first 3 modes at ~146, ~208, ~223 cm$^{-1}$ (defined as $A_{1g}^1$ + $B_{1g}^1$, $A_g^2$, and $B_{1g}^2$) are dominated by the movements of Se atoms, while the highest modes at 260 cm$^{-1}$ and ~270 cm$^{-1}$ (defined as $A_{1g}^3$ and $A_{1g}^3$) involve the relative movements between Pd and Se atoms. The measured peaks are slightly shifted with respect to the four main modes of bulk PdSe$_2$ (at 143, 206, 222 and 256 cm$^{-1}$), confirming the few layer nature of the flake.[14,26]

The device electric measurements were carried out using a Keithley 4200 semiconductor analyzer in a two-terminal configuration. The samples were measured inside a SEM chamber (ZEISS, LEO 1530) at room temperature, in dark and, if not otherwise stated, with controlled pressure below $10^{-6}$ Torr. Source and drain (C1-C2 or C2-C3 leads in Figure 1(b)) were contacted with piezoelectric-driven tungsten probes while the sample holder electrically connected to the Si substrate by silver paint worked as the gate terminal.

Drain biases resulting in a current higher than 1 µA or $|V_{gs}|$ > 50 V were avoided to prevent channel or gate dielectric damages.

Most of the measurement discussed in the paper were referred to the transistor formed between leads C1 and C2 (Figure 1(b)) characterized by channel length L=2.0 µm and width W=4.75 µm; the similar transistor, formed by leads C2 and C3, gave comparable results and was used as confirmation.

**Acknowledgements**

We acknowledge the economic support of POR Campania FSE 2014–2020, Asse III Ob. specifico l4, D.D. n. 80, 31/05/2016 and CNR-SPIN SEED Project 2017 and Project PICO & PRO, ARS S01_01061, PON "Ricerca e Innovazione" 2014-2020. Feng Miao would like to acknowledge the support by National Key Basic Research Program of China (2015CB921600), the National Natural Science Foundation of China (61625402, 61574076), and the Collaborative Innovation Center of Advanced Microstructures. Shi-Jun Liang would like to acknowledge the support of Natural Science Foundation of Jiangsu Province (BK20180330).

**Conflict of Interest**

The authors declare no conflict of interest.